\documentstyle[prl,aps,twocolumn,floats,amssymb,epsfig]{revtex}

\newcommand{\mb}[1]{\ifmmode#1\else\mbox{$#1$}\fi}

\newcommand\de{\mb{\delta}}

\newcommand\calD{\mb{{\cal D}}}

\newcommand\calH{\mb{{\cal H}}}

\newcommand\calK{\mb{{\cal K}}}
\newcommand\calL{\mb{{\cal L}}}

\newcommand\calV{\mb{{\cal V}}}

\newcommand{\beq}{\begin{equation}}
\newcommand{\eeq}{\end{equation}}
\newcommand{\nn}{\nonumber}
\newcommand{\bea}{\begin{eqnarray}}
\newcommand{\eea}{\end{eqnarray}}

\newcommand{\gsim}
{\raise.3ex\hbox{$\;>$\kern-.75em\lower1ex\hbox{$\sim$}$\:$}}
\newcommand{\lsim}
{\raise.3ex\hbox{$\;<$\kern-.75em\lower1ex\hbox{$\sim$}$\:$}}
\newcommand{\ts}{\textstyle}

\newcommand{\half}{{\ts \frac{1}{2}}}

\newcommand{\qu}{{\rm qu}}

\begin{document}
\renewcommand{\thesubsection}{\arabic{subsection}}
\draft


\twocolumn[\hsize\textwidth\columnwidth\hsize\csname @twocolumnfalse\endcsname
\author{Nathan\  F.\  Lepora}
\title{A comment on: `Some problems with calculating the quantum corrections 
to the classical 't~Hooft-Polyakov monopole'} 
\date{June 22, 2003}
\maketitle

\begin{abstract}
In a recent publication we noticed that the Hamiltonian density for
fluctuations around the 't~Hooft-Polyakov monopole appeared to be
non-Hermitian. Here we show that when this Hamiltonian density is 
integrated into the Hamiltonian all non-Hermitian terms
give a vanishing total contribution.\\
\end{abstract}]


In a recent letter~\cite{qu} we were looking at
the quantum fluctuations around the 't~Hooft-Polyakov monopole
and noticed that the Hamiltonian density appeared to be
non-Hermitian. We then speculated that this could indicate a subtlety in the 
behaviour of monopoles, although we also said the
problem may be merely technical.

The aim of this comment is to show that there are no problems with
the Hermiticity of the Hamiltonian. When the Hamiltonian
density is integrated to the Hamiltonian the non-Hermitian 
terms give a vanishing contribution.

For illustration we consider the Hamiltonian density for the gauge
fluctuations around the monopole
\bea
\label{calH}
\calH_\qu=\half\pi^i_a\pi^i_a+\half a^i_a\left(-\calK^{ij}_{ab}
+\calV^{ij}_{ab}+\calD^{ij}_{ab}\right)a^j_b,
\eea
where $\calV^{i}_{ab}=\calV^{ji}_{ba}$, 
$\calK^{ij}_{ab}=\de_{ab}\left(\nabla^2\de^{ij}-\partial^i\partial^j\right)$
and $\calD^{ij}_{ab}=D^{ij}_{abl}\partial_l$.
The conjugate momenta are $\pi^i_a=\dot a^i_a$ in a temporal gauge.
Then the first few terms in this Hamiltonian are Hermitian because
\bea
\int\!\!d^3r\,a^i_a\calK^{ij}_{ab} b^j_b =
\int\!\!d^3r\,b^i_a\calK^{ij}_{ab} a^j_b,\ \ 
a^i_a\calV^{ij}_{ab}b^j_b=b^i_a\calV^{ij}_{ab}a^j_b;
\eea
the first by partial integration, neglecting surface terms, and
the second by $\calV^{ij}_{ab}=\calV^{ji}_{ba}$.
However, a partial integration of $\calD^{ij}_{ab}=D^{ij}_{abl}\partial_l$
implies
\beq
\label{ddag}
(\calD^\dagger)^{ij}_{ab}=
-\calD^{ji}_{ba}-\partial_lD^{ji}_{bal}\neq
\calD^{ij}_{ab},
\eeq
whereby $\calD^{ij}_{ab}$ is not Hermitian, as pointed out in ref.~\cite{qu}.

This non-Hermiticity is misleading though. Although the Hamiltonian density
$\calH_\qu$ appears to contain non-Hermitian terms,
both the Hamiltonian $H_\qu=\int\! d^3r\,\calH_\qu$ and the field equations
contain only Hermitian operators. 

Let us consider the Hamiltonian and the field equations separately:

\noindent (i) Notice that
$\calD^{ij}_{ab}$ in (\ref{calH}) can be split 
into Hermitian and anti-Hermitian components:
\bea
\label{arg1}
\calD=\calD_H+\calD_A,\hspace{1em}\calD_H^\dagger=\calD_H,\ 
\calD_A^\dagger=-\calD_A,
\eea
defined by $\calD_H=\half(\calD+\calD^\dagger)$,
$\calD_A=\half(\calD-\calD^\dagger)$. Now, the Hamiltonian can
be written as (the dots denote other terms in the
Hamiltonian)
\bea
\label{arg2}
H_\qu[a^i_a,\pi^i_a]&=&\int\!d^3r\,\calH_\qu[a^i_a,\pi^i_a]\nn\\
&=&\cdots + \int\!d^3r\, a^i_a(\calD_H)^{ij}_{ab}a^j_b
+\int\!d^3r\,(\calD_A)^{ij}_{ab}a^j_b\nn\\
&=&\cdots + \int\!d^3r\, a^i_a(\calD_H)^{ij}_{ab}a^j_b
\eea
because of
$\int d^3r\,a^i_a(\calD_A)^{ij}_{ab}a^j_b=
-\int d^3r\,a^i_a(\calD_A)^{ij}_{ab} a^j_b = 0$.
Thus the non-Hermitian part $\calD_A$ does not contribute to the Hamiltonian, 
which then contains only Hermitian terms.

\noindent (ii) The second argument uses the field
equations~\cite{comm}. Consider, for instance, the Lagrangian 
density for gauge fluctuations in a temporal $a^0_a=0$ gauge,
\bea
\label{arg3}
\calL_\qu[a^i_a]&=&\half\dot a^i_a\dot a^i_a+
\half\partial^ia^j_a\partial^ia^j_a-\half\partial^ia^j_a\partial^ja^i_a\nn\\
&&\hspace{4em}+\,\half a^i_a\calV^{ij}_{ab}a^j_b
+ \half a^i_aD^{ij}_{abl}\partial_la^j_b.
\eea
One must also impose an additional Gauss's law constraint from this choice
of gauge. From (\ref{arg3}) the field equations are
\bea
-\ddot a^i_a+\calK^{ij}_{ab}a^j_b+
\half(\calV^{ij}_{ab}+\calV^{ji}_{ba})a^j_b
\hspace{7em}\nn\\
+\,\half\calD^{ij}_{ab}a^j_b 
+\half(-\calD^{ji}_{ba}-\partial_lD^{ji}_{bal})a^j_b=0.
\eea
By (\ref{ddag}) the last operator is recognized as 
$(\calD^\dagger)^{ij}_{ab}$ and the field equations are
\beq
\label{arg4}
-\ddot a^i_a+
\left(\calK^{ij}_{ab}+\calV^{ij}_{ab}+(\calD_H)^{ij}_{ab}\right)a^j_b=0.
\eeq
Again, only the Hermitian component contributes to the equations of motion.

Finally, we note it is not surprising that the operators in the 
Hamiltonian $H_\qu$ are Hermitian. The quadratic-order Hamiltonian also
describes the scattering of scalar and gauge bosons off the classical
monopole background. For the $S$-matrix to be unitary one
should expect a Hermitian Hamiltonian.


\end{document}